\documentclass[manuscript]{acmart}

\usepackage[ruled,vlined,linesnumbered]{algorithm2e}
\usepackage{arydshln}
\usepackage{amsmath}

\AtBeginDocument{%
  \providecommand\BibTeX{{%
    \normalfont B\kern-0.5em{\scshape i\kern-0.25em b}\kern-0.8em\TeX}}}

\usepackage{blindtext}
\usepackage{tcolorbox}
\usepackage{graphicx}
\usepackage{xspace}
\usepackage{multirow}

\newcommand{\systemname}{\textsc{ILuvUI}\xspace}

\setcopyright{acmcopyright}
\copyrightyear{2023}
\acmYear{2023}

\acmSubmissionID{}

\begin{document}

\title{\textsc{ILuvUI}: Instruction-tuned LangUage-Vision modeling of UIs from Machine Conversations}
\newcommand{\name}[1]{\def\papername{#1}}
\name{ILuvUI}

\author{Yue Jiang}
\affiliation{%
  \institution{Aalto University}
  \city{Espoo}
  \country{Finland}
}

\author{Eldon Schoop}
\affiliation{%
  \institution{Apple}
  \city{Seattle}
  \country{USA}
}

\author{Amanda Swearngin}
\affiliation{%
  \institution{Apple}
  \city{Seattle}
  \country{USA}
}

\author{Jeffrey Nichols}
\affiliation{%
  \institution{Apple}
  \city{Seattle}
  \country{USA}
}

\renewcommand{\shortauthors}{Trovato and Tobin, et al.}

\begin{abstract}
Multimodal Vision-Language Models (VLMs) enable powerful applications from their fused understanding of images and language, but many perform poorly on UI tasks due to the lack of UI training data. In this paper, we adapt a recipe for generating paired text-image training data for VLMs to the UI domain by combining existing pixel-based methods with a Large Language Model (LLM). Unlike prior art, our method requires no human-provided annotations, and it can be applied to any dataset of UI screenshots. We generate a dataset of 335K conversational examples paired with UIs that cover Q\&A, UI descriptions, and planning, and use it to fine-tune a conversational VLM for UI tasks. To assess the performance of our model, we benchmark it on UI element detection tasks, evaluate response quality, and showcase its applicability to multi-step UI navigation and planning.
\end{abstract}

\ccsdesc[500]{Human-centered computing~Interactive systems and tools}
\ccsdesc[300]{Human-centered computing~Natural language interfaces}
\ccsdesc[300]{Computing methodologies~Scene understanding}

\ccsdesc[500]{Human-centered computing~User interface toolkits}

\newcommand\DELETE[1]{\textcolor{red}{#1}}
\newcommand\ADD[1]{\textcolor{black}{#1}}
\newenvironment{delete}{\par\color{red}}{\par}
\newenvironment{add}{\par\color{black}}{\par}
\newcommand{\loss}{\mathcal{L}}
\newcommand{\image}{\mathcal{I}}
\newcommand{\encoder}{\mathcal{E}}
\newcommand{\decoder}{\mathcal{D}}
\newcommand{\normal}{\mathcal{N}}

\keywords{}

\begin{teaserfigure}
\centering
\includegraphics[width=0.8\textwidth]{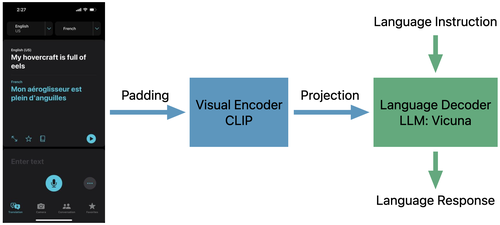}
  \caption{We generate a dataset of Q\&A pairs that cover several UI description and reasoning tasks and use it to fine-tune a Vision-Language Model (VLM) that accepts multimodal inputs from screenshot pixels and a user-provided text input. Our model, \systemname, enables new kinds of UI tasks with conversational VLMs.}
  \Description{}
  \label{fig:model}
 \end{teaserfigure}

\maketitle

\section{Introduction}

For nearly as long as graphical user interfaces (UIs) have been popular, users have sought methods to verbally describe them for accessibility~\cite{edwards1995screenreader} or automate interaction with them, for example to execute repetitive tasks~\cite{cypher1991eager}.
Understanding and automating actions on UIs is a challenging task since the UI elements in a screen, such as list items, checkboxes, and text fields, encode many layers of information beyond their affordances for interactivity alone. The layout, visual style, and textual content of a UI are often designed to afford user expectations for the domain and capabilities of the application, and the elements themselves may be dynamic or stateful.
This inherent complexity makes comprehending and conveying UI-related information through natural language challenging.

Many works in HCI have sought to automate tasks on UIs using programming by demonstration~\cite{cypher1991eager, coscripter, sugilite}, mining similar screenshots through past interactions~\cite{savant}, and Large Language Models (LLMs)~\cite{bryanConversational, chunyangChattingTesting, tobyEmpoweringLLM}.
LLMs in particular have demonstrated remarkable abilities to comprehend task instructions in natural language in many domains~\cite{instructgpt, peng2023instruction, chiang2023vicuna, touvron2023llama, zheng2023judging}, 
however using text descriptions of UIs alone with LLMs leaves out the rich visual information of the UI.
Fusing visual with textual information is important to understanding UIs as it mirrors how many humans engage with the world.
One approach that has sought to bridge this gap when applied to natural images are Vision-Language Models (VLMs), which accept \emph{multimodal} inputs of both images and text, typically output only text, and allow for general-purpose question answering, visual reasoning, scene descriptions, and conversations with image inputs~\cite{instructblip, liu2023visual, chen2023pali, alayrac2022flamingo}.
However, the performance of these models on UI tasks fall short compared to natural images because of the lack of UI examples in their training data.

In this paper, we adapt the LLaVA~\cite{liu2023visual} VLM training data generation recipe to the UI domain. Our data generation recipe does not require any human labeling, and can be adapted to UIs in existing datasets such as Rico~\cite{deka2017rico}, which do not have existing textual descriptions.
We generate a dataset of 335K image-instruction pairs using screenshots from the AMP dataset~\cite{zhang2021screen} and additional data from the Crawls interaction trace dataset~\cite{feiz2022understanding}. In order to generate text pairs for a given screenshot, we employ a UI element detection model~\cite{zhang2021screen} that identifies the elements in a screen, converts these detections into a structured text-based representation, and then prompts an LLM with this representation and additional context to generate one or more realistic phrases. We generate six types of phrases using different prompts: single-step Q\&A conversations, detailed descriptions, listing available actions, predicting UI action outcomes, selecting a UI element given a goal, and a goal-based plan.

After generating data using the gpt3.5-turbo LLM~\cite{chatgpt}, we use that data to fine-tune a conversational VLM. Our resulting model, \systemname, is capable of describing properties of UI elements and screens, contextual help, and planning multi-step interactions.
Like many large, unsupervised models, \systemname is capable of tasks it was not trained to perform. Beyond interpreting the complexities of UIs and following instructions, our model paves the way towards using VLMs to enhance UI accessibility by automatically generating descriptions and acting on instructions provided by speech.
To better understand \systemname's performance, we benchmark against a UI element detection model, evaluate its response quality compared to a baseline VLM, and show selected examples which demonstrate \systemname's planning and reasoning capabilities.

Our paper makes the following contributions:
\begin{enumerate}
    \item We adapt the LLaVA method~\cite{liu2023visual} to generate paired text-image data to train VLMs in the UI domain using an LLM and a UI element detector. Our method only uses UI screenshots as input, does not require any human-provided captions, and produces six different types of text pairs. 
    \item We fine-tune an open-source VLM, LLaVA~\cite{liu2023visual}, on our UI dataset, leading to the development of a UI-focused instruction-following visual agent, and assess its performance with several lightweight evaluations.
\end{enumerate}
\section{Related Work}

\subsection{UI Understanding}

Pixel-based UI understanding is broadly applicable across diverse domains, including interface adaptation~\cite{banovic2012waken, chang2011associating, dixon2010prefab, zettlemoyer1999visual}, GUI testing~\cite{yeh2009sikuli}, data-driven GUI searches~\cite{chen2019gallery, chen2020wireframe, huang2019swire}, prototyping~\cite{swearngin2018rewire}, UI code generation to support app development~\cite{beltramelli2018Pix2code, chen2018from, chen2020object, de2021using, nguyen2015reverse}, and GUI security~\cite{chen2021gui}.
To identify UI elements within images, conventional image processing methods have long been employed, often relying on detecting and aggregating edges~\cite{nguyen2015reverse, swearngin2018rewire, yongxin2019ui2code}. While adept at handling simpler GUIs, these methodologies struggle when faced with images that have gradients, photographs, or intricate UI layouts. Furthermore, template matching techniques~\cite{dixon2010prefab, pongnumkul2011pause, yeh2009sikuli, zettlemoyer1999visual} require meticulous feature engineering and curated templates, introducing potential limitations when applied to UIs with diverse visual attributes. 
Many reverse engineering approaches also predict UI elements and layouts~\cite{yeh2009sikuli, nguyen2015reverse, dixon2010prefab, dixon2011content, dixon2014prefab, christof2008automated, Katsimpa2006application, Bernardi2009REUWA, Ascar2013Model, jiang2019orclayout, jiang2020orcsolver, jiang2021reverseorc, swearngin2017genie, bielik2018robust, krosnick2018expresso, moore1996rule}, detect hierarchical groupings~\cite{wu2021screen}, generate code for UIs~\cite{beltramelli2018Pix2code, yongxin2019ui2code}, and enable platform migration or UI improvement~\cite{moore1997using, moore1998user, STROULIA2003User, Lucca2004Reverse}. 

The surge of deep learning-based approaches has been enabled by comprehensive UI datasets such as Rico~\cite{deka2017rico}. A prominent instance of this trend is witnessed in pix2Code~\cite{beltramelli2018Pix2code}, which employs an end-to-end neural image captioning model to generate descriptions of interface layouts. Correspondingly, Chen et al.~\cite{chen2018from} uses a CNN-RNN model to create a UI skeleton that includes widget types and layouts inferred from screenshots. 
Other models~\cite{chen2019gallery, white2019improving} use object detection techniques to detect GUI widgets within screenshots. Some prior research~\cite{schoop2022predicting, swearngin2019modeling} predicts the tappability of mobile app UI elements. Additionally, some deep learning models use crowdsourcing to generate image captions~\cite{gleason2020twitter, guinness2018caption}.
Other approaches~\cite{chen2020object, chen2019automated, moran2020machine} combine traditional image processing methods, such as edge detection, with deep learning-based classification models to detect UI elements and semantics, including UI types. Zhang et al.~\cite{zhang2021screen} infers interface elements and metadata from pixels, including UI types, navigation order, groupings, image descriptions, and icon classes. We make use of this latter method to produce textual representations of UIs that we use to generate text-screenshot pairs for VLM training.

\subsection{Large Language Models and Vision Language Models}
Large language models (LLMs) have broad potential in interpreting language, serving as an interface for a versatile AI assistant, are capable of comprehending explicit task instructions in natural language. ChatGPT~\cite{chatgpt} and GPT-4~\cite{gpt4} show the proficiency of well-aligned LLMs in following human instructions. This achievement has sparked a surge of interest in the development of open-source LLMs.
Among these, LLaMA~\cite{touvron2023llama} is a downloadable LLM that rivals the performance of GPT-3. Other models like Alpaca~\cite{alpaca}, Vicuna~\cite{chiang2023vicuna}, and GPT-4-LLM~\cite{gpt4} have leveraged machine-generated, high-quality instruction-following examples to enhance the alignment capabilities of LLMs. The LLaVA~\cite{liu2023visual} model on which we base our work is itself a combination of the CLIP~\cite{clip} and Vicuna~\cite{chiang2023vicuna} models.

Spotlight is an example of a VLM that has been trained for UI tasks~\cite{Li2022SpotlightMU}. This model takes as input both an image of UI and a region of interest, and outputs text related to the region, and was applied to tasks such as widget captioning, screen summarization, command grounding, and tappability prediction. The Spotlight model is quite powerful, however it is somewhat limited by the requirement that a region of interest be specified, which may not knowable in advance for UI automation tasks without the use of other technologies. \systemname in contrast does not require a region of interest, and accepts a text prompt as input in addition to the UI image, which enables it to provide answers for use cases such as visual question answering.

\section{Generating Multimodal Data for User Interfaces}

\begin{figure}[t]
\centering
\includegraphics[width=0.8\textwidth]{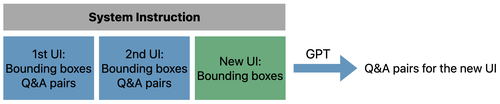}
  \caption{We use two author-created ``golden examples'' of UI elements and Q\&A pairs as few-show examples for data generation. These examples are prepended to a third list of UI elements, where gpt3.5-turbo predicts new, resulting Q\&A pairs.}
  \Description{}
  \label{fig:few_shot}
 \end{figure}

While general-purpose VLMs perform well on natural images for several tasks, their capabilities are limited when performing tasks with UIs. This is because existing VLMs lack latent knowledge of common UI design principles, tasks, and element types.
Several datasets exist in the literature that pair natural images with text descriptions, Q\&A, or instructions, but the availability of training data that ties UI screenshots with descriptions or grounded conversation is comparatively limited.
This motivates the need to create a dataset that pairs UIs with natural language instructions and descriptions to enable training a VLM that can understand and act on UIs.

LLaVA~\cite{liu2023visual} is recent work that introduces a method for creating text-image data pairs from an existing image dataset and a multimodal VLM trained on that data which connects the CLIP vision encoder~\cite{clip} with the Vicuna LLM~\cite{chiang2023vicuna} for general-purpose visual and language understanding. We adapt both aspects of this work to create a VLM tuned to UI tasks.

A key contribution of LLaVA's data generation method is using an off-the-shelf LLM to generate a dataset of text-image pairs from the existing COCO image dataset~\cite{coco}. The COCO dataset pairs images with human-annotated captions and bounding boxes of objects within each image. For each image, the method combines the captions, list of bounding boxes of objects with the object types, and an additional prompt into a request to a proprietary LLM (experiments were done with both GPT3.5 and GPT4). The resulting outputs from the LLM contained realistic text phrases that were paired with the original image to produce a VLM training set. LLaVA produced three kinds of text pairs that they termed Q\&A, detailed description, and complex reasoning.

\begin{figure*}
\noindent\begin{minipage}{\textwidth}
\begin{tcolorbox}[colback=brown!3!white,colframe=brown!50!black,
  colbacktitle=cyan!75!black]
  \begin{minipage}[t]{0.11\linewidth}
    \vspace*{6pt}
    \textcolor{blue}{\textbf{Input UI:}} 
    \end{minipage}%
\begin{minipage}[t]{0.32\linewidth}
    \vspace*{0pt}
        \includegraphics[width=0.6\linewidth]{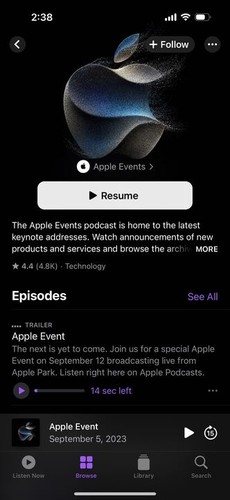}
    \end{minipage} %
     \begin{minipage}[t]{0.05\linewidth}
    \vspace*{8pt}
    \textcolor{blue}{\textbf{}} 
    \end{minipage}%
    \begin{minipage}[t]{0.15\linewidth}
    \vspace*{6pt}
    \textcolor{blue}{\textbf{UI Detection:}} 
    \end{minipage}%
    \begin{minipage}[t]{0.32\linewidth}
    \vspace*{0pt}
        \includegraphics[width=0.6\linewidth]{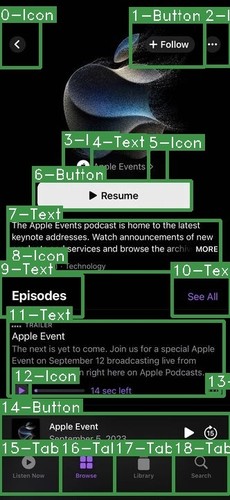}
    \end{minipage} %
  \tcblower
  \textcolor{blue}{\textbf{Formatted Bounding Boxes:}} \\
\small{Label: Icon (Type: back), BoundingBox from (0, 30) to (58, 97)} \\
\small{Label: Button, Text: add, Follow, BoundingBox from (188, 31) to (295, 96)} \\
\small{Label: Icon (Type: more), BoundingBox from (295, 31) to (332, 95)} \\
\small{Label: Icon (Type: refresh), BoundingBox from (91, 210) to (132, 260)} \\
\small{Label: Text, Text: Apple Events, BoundingBox from (132, 213) to (214, 260)} \\
\small{Label: Icon (Type: right arrow), BoundingBox from (214, 214) to (241, 260)} \\ 
\small{Label: Button, Text: Resume, BoundingBox from (48, 260) to (276, 302)} \\
\small{Label: Text, Text: The Apple Events podcast is home to the latest, keynote addresses. Watch announcements of new, products and services and browse the archiv MORE, 4.4 (4.8K) Technology, BoundingBox from (11, 317) to (318, 391)} \\
\small{Label: Icon (Type: star), BoundingBox from (16, 377) to (26, 388)} \\
\small{Label: Text, Text: Episodes, BoundingBox from (0, 396) to (119, 456)} \\
\small{Label: Text, Text: See All, BoundingBox from (248, 398) to (332, 455)} \\
\small{Label: Text, Text: The next is yet to come. Join US for a special Apple, Event on September 12 broadcasting live from, Apple Park. Listen right here on Apple Podcasts., .... TRAILER, 14 sec left, BoundingBox from (11, 462) to (323, 570)} \\
\small{Label: Icon (Type: play), BoundingBox from (19, 550) to (43, 573)} \\
\small{Label: Icon (Type: more), BoundingBox from (298, 558) to (317, 566)} \\
\small{Label: Button, Text: September 5, 2023, Apple Event, play, BoundingBox from (0, 595) to (332, 649)} \\
\small{Label: Tab, Text: Listen Now, BoundingBox from (0, 654) to (87, 711)} \\
\small{Label: Tab, Text: Browse, BoundingBox from (87, 654) to (165, 711)} \\
\small{Label: Tab, Text: Library, BoundingBox from (165, 654) to (250, 711)} \\
\small{Label: Tab, Text: Search, BoundingBox from (250, 654) to (330, 711)} \\
\textcolor{blue}{\textbf{Caption:}} \\
This UI screen displays an Apple Events podcast page, allowing users to browse and listen to episodes, view upcoming events, and navigate between different tabs for listening, browsing, library, and search.
\end{tcolorbox}
\vspace{-6mm}
\captionof{figure}{One example to illustrate the input UI image, the UI detection results, formatted bounding boxes and caption.  }\label{ui_input}
\end{minipage}
\end{figure*}

We adapted the LLaVA dataset generation recipe to UI tasks in several ways, to fit our existing UI dataset and different UI tasks. The LLaVA method relied on the COCO dataset~\cite{coco} to provide both human-annotated captions and object bounding boxes. For UI data, we use the AMP dataset  from Zhang et al.~\cite{zhang2021screen}, which comprises 80,945 unique screens from 4,239 iPhone apps; and the \textsc{Crawls} dataset~\cite{feiz2022understanding}, which contains 750,000 iOS app screens from 6,000 apps. These datasets contain screenshots and annotations of UI element bounding boxes, but no human-annotated captions of its UIs. Ultimately and although they are of high quality, we decided not to use the UI element bounding box annotations, and instead produced the bounding boxes using an object detection model~\cite{zhang2021screen} during data generation as we sought a method that would work with completely unannotated UI examples as might be found in other datasets. The missing UI captions were generated using an LLM for each screen with a prompt that included the UI element detections.  We then created text pairs by generating analogous examples to the original recipe (Q\&A, detailed descriptions), as well as adapting LLaVA's original complex reasoning task to multiple different UI tasks: listing possible actions, predicting the outcome of actions, selecting an element that accomplishes a goal, and goal-based planning. \autoref{ui_input} provides an example of this process.

In total, we generate 353K unique language-image instruction-following samples, including 224K in conversations, 32K in concise description, 32K in detailed description, 32K in logical reasoning, 32K in potential actions, and 1K in UI transition, respectively.

\subsection{Detecting UI Elements and Generating Screen Captions}

For each screen, we use the UI detection model from Zhang, et al~\cite{zhang2021screen} to generate a list of UI elements, their bounding boxes and types. UI element information is included in all of our data generation prompts and is formatted as follows:

\begin{center}
    \texttt{Label: [type], Text: [text], BoundingBox from (x1, y1) to (x2, y2)}
\end{center}

\texttt{[type]} refers to the element category (e.g., text, button, or icon), \texttt{[text]} is the text contained by the UI element extracted via OCR, and the bounding box coordinates are defined by the top-left position and the bottom-right position of the UI element, respectively.

In order to generate screen captions, which are also used in the rest of our data generation prompts, we query gpt-3.5-turbo with formatted the detected UI elements using the following prompt: ``Given the UI screen [Bounding Boxes]. Write a single-sentence usage description for this UI screen.'' This prompt was determined experimentally and seemed to produce good results.

\subsection{Generating LLaVA-Style Data}

LLaVA generates both single-step Q\&A conversations and detailed descriptions as part of its data generation procedure, which we also include in our own procedure. For both, we use the same few-shot in-context learning method as LLaVA to present gpt-3.5 with examples of desired output using 2 ``golden examples'' that we authored. As shown in \autoref{fig:few_shot}, we design a system message and select two UI examples with their respective bounding boxes, captions, and desired question and responses for each data type.
Example prompts for each data type are provided in Supplementary Material.

\paragraph{Single-Step Q\&A Conversations}
The Q\&A data type consists of question-answer pairs between an assistant and a user, conditioned on a source UI screenshot. The assistant responds in a way that suggests they are viewing the UI image while answering the question. A variety of questions are asked about the UI image, including element attributes, placements, relative positions, functionalities, and purpose. Only questions with definite answers are considered.

\paragraph{Detailed Description}
The Detailed Description data type is a rich and comprehensive description of a UI image describing all of its elements and their respective functionality. We elicit descriptions from a list of various questions that ask the LLM to describe the UI in detail and in terms of the elements on that screen. For each UI screenshot example, we randomly select a single question from this list to generate the description.

\subsection{Generating UI-Specific Data}

LLaVA includes a generic ``complex reasoning'' task as part of its data generation process, which elicited complex responses including common-sense or historical descriptions of elements in the image, and step-by-step descriptions of actions taking place in a scene. We found that this task was too general for UI tasks, and instead chose to substitute it with four UI-specific data generation types: identifying possible actions in a screen, predicting the outcome of taking an action, selecting a UI element capable of a given action, and formulating a plan that accomplishes a goal on the given screen. These tasks were designed to familiarize the VLM with the capabilities of particular UI components, and to convey a sense of the affordances that are signified by common UI design patterns~\cite{liao2022rediscoveraffordance}.

\subsubsection{Listing Available UI Actions}
For our first task, gpt-3.5-turbo is prompted to list all of the potential actions that can be taken on a particular UI, given its high level description and the list of recognized elements on the screen. Actions can range from tapping, swiping, or entering text depending on the functionality of the UI.
For this task, we construct a prompt from the UI screenshot caption and its bounding boxes, and use the zero-shot result from gpt-3.5-turbo.

\subsubsection{Predicting the Outcome of UI Actions}
In addition to understanding the capabilities of various UI elements, we wanted our VLM to be able to reason about the potential outcomes of acting on UI elements.
From the interaction traces in the Crawls dataset~\cite{feiz2022understanding}, which overlaps with the AMP dataset, we were able to establish ground truth correspondences from an element that was tapped on one screen and the second screen to which that interaction led.
This dataset includes the pixel coordinates of a tap action that caused the UI transition. We run UI element detection on the first screen and match the pixel to the UI element with the smallest bounding box that contains the tap location to identify what specific element was tapped to cause the transition.

We feed the UI elements of the first screen, tapped UI element, and caption of the second screen to gpt-3.5-turbo, and prompt it to formulate a single QA pair that asks what action will take place when the given UI element is tapped, and answer with a concise description of the second screen.
Since the LLM has access to ground truth in its context for these cases, there is no need to use few-shot examples.

\subsubsection{Selecting a UI Element Given a Goal}
As a complement to the above task, we also use the LLM to generate a QA pair that asks what element must be tapped on the first screen in order to arrive at a target view, specified by an concise description of the second screen. The question is prompted to not include any information about the second screen.
Similar to the above, this QA pair also uses Crawls dataset, so there is no need to use few-shot since ground truth is available.

\subsubsection{Goal-Based Planning}
The last UI task we use for data generation is goal-based planning, where a QA pair is generated to ask for a directions to accomplish a task in the specified UI.
Unlike the pairs generated with the Crawls dataset, the generated directions need not result in opening a new view. Generated directions can interact with the current view (e.g., to change its state), formulate a multi-step plan across multiple views, or relate a task with the current UI to commonsense knowledge.
In practice, most generated examples formulated multi-step plans for completing complex tasks on the given screen.
For this task, we use two 2-shot examples. 
The first example asks what step should be taken to log into a given login screen (first, enter email; then password; then click ``sign in''). The second example asks what the user should do when the item shown on a product page is too expensive (open a visible details menu to see if there are promotions or discounts available).

\newpage

\begin{figure*}
\noindent\begin{minipage}{\textwidth}
\begin{tcolorbox}[colback=blue!3!white,colframe=blue!50!black,
  colbacktitle=blue!75!black]
\textcolor{blue}{\textbf{Conversation:}} \\
Question: What are the options available in the navigation bar at the bottom?\\
Answer: The options available in the navigation bar at the bottom are ``Listen Now'', ``Browse'', ``Library'', and ``Search''.\\
Question: What does the "Follow" button do?\\
Answer: The ``Follow'' button allows users to follow the Apple Events podcast, which means they will receive updates or notifications when new episodes are released. \\
Question: What information is provided about the Apple Events podcast? \\
Answer: The information provided about the Apple Events podcast includes a description of the podcast, its rating, and its category. The description mentions that the podcast is home to the latest keynote addresses, and users can watch announcements of new products and services and browse the archive. \\
Question: Where is the ``Back'' icon? \\
Answer: The ``Back'' icon is at the top-left corner.\\ \\
\textcolor{blue}{\textbf{Detailed Description:}} \\
Question: Describe the image in detail.\\
Answer: \\
This UI screen presents an Apple Events podcast page. At the top, there are three icons: a back button on the left, a 'Follow' button in the middle, and a 'more' icon on the right. Just below these, a refresh icon is positioned on the left, followed by the title 'Apple Events' and a right arrow icon. A 'Resume' button is situated beneath these elements.

The page includes a description of the Apple Events podcast, which reads: "The Apple Events podcast is home to the latest, keynote addresses. Watch announcements of new, products and services and browse the archive MORE, 4.4 (4.8K) Technology". A star icon is present within this text block, likely indicating a rating or favorite feature.

The 'Episodes' section is found below the description, with a 'See All' option on the right. The first episode listed is titled "The next is yet to come. Join US for a special Apple, Event on September 12 broadcasting live from, Apple Park. Listen right here on Apple Podcasts., .... TRAILER, 14 sec left". This episode entry includes a play icon on the left and a 'more' icon on the right. 

Below this episode, there's a button labeled "September 5, 2023, Apple Event, play", which likely represents an upcoming event. 

At the bottom of the screen, there are four tabs for navigation: 'Listen Now', 'Browse', 'Library', and 'Search', listed from left to right. These tabs allow users to switch between different sections of the podcast page. 
\end{tcolorbox}
\captionof{figure}{One example to illustrate the data showing the four types of responses for UI understanding.}\label{response}
\end{minipage}
\end{figure*}

\newpage

\begin{figure*}
\noindent\begin{minipage}{\textwidth}
\begin{tcolorbox}[colback=blue!3!white,colframe=blue!50!black,
  colbacktitle=blue!75!black]
\textcolor{blue}{\textbf{Listing Available UI Actions}} \\
Question: What actions can we perform on this user interface? \\
Answer: \\
- Tap on the "Button" with the text "add" \\
- Tap on the "Button" with the text "Resume" \\
- Tap on the "Icon" (Type: star) \\
- Tap on the "Text" with the text "See All" \\
- Tap on the "Icon" (Type: play) \\
- Tap on the "Icon" (Type: more) \\
- Tap on the "Button" with the text "September 5, 2023, Apple Event, play" \\
- Tap on the "Tab" with the text "Listen Now" \\
- Tap on the "Tab" with the text "Browse" \\
- Tap on the "Tab" with the text "Library" \\
- Tap on the "Tab" with the text "Search" \\
\\
\textcolor{blue}{\textbf{Goal-Based Planning}} \\
Question: What should the user do if they want to listen to the latest Apple Event podcast episode? \\
Answer: \\
To listen to the latest Apple Event podcast episode, the user should look under the "Episodes" section, which is located towards the middle of the UI screen. The most recent episode is typically listed first. The user can identify the episode by its title, which in this case is "Apple Event". To start listening, the user should click on the "play" icon located within the bounding box of the episode. \\ 
\end{tcolorbox}
\captionof{figure}{One example to illustrate the data showing the four types of responses for UI understanding.}\label{response}
\end{minipage}
\end{figure*}

\newpage

\begin{figure*}[!]
\noindent\begin{minipage}{\textwidth}
\begin{tcolorbox}[colback=cyan!3!white,colframe=cyan!50!black,
  colbacktitle=yellow!75!black]
  \begin{minipage}[t]{0.125\linewidth}
    \vspace*{6pt}
    \textcolor{blue}{\textbf{Current UI:}} 
    \end{minipage}%
\begin{minipage}[t]{0.32\linewidth}
    \vspace*{0pt}
        \includegraphics[width=0.5\linewidth]{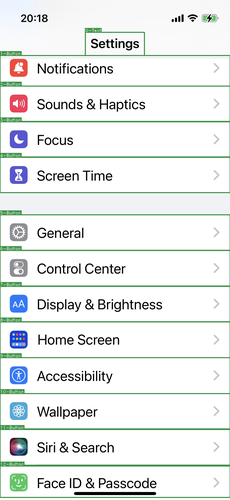}
    \end{minipage} %
     \begin{minipage}[t]{0.03\linewidth}
    \vspace*{8pt}
    \textcolor{blue}{\textbf{}} 
    \end{minipage}%
    \begin{minipage}[t]{0.16\linewidth}
    \vspace*{6pt}
    \textcolor{blue}{\textbf{Subsequent UI:}} 
    \end{minipage}%
    \begin{minipage}[t]{0.32\linewidth}
    \vspace*{0pt}
        \includegraphics[width=0.5\linewidth]{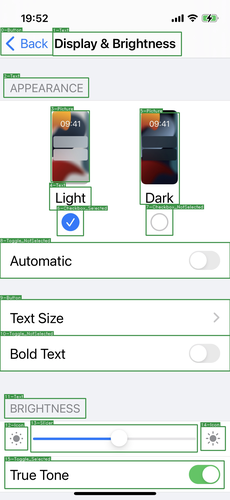}
    \end{minipage} \\ \\ %
  \textcolor{blue}{\textbf{Formatted Bounding Boxes of the Current UI:}} \\
Label: Text, Text: Settings, BoundingBox from (417, 157) to (706, 270)\\
Label: Button, Text: notification, Notifications, right arrow, BoundingBox from (0, 270) to (1125, 419)\\
Label: Button, Text: speaker, Sounds \& Haptics, right arrow, BoundingBox from (0, 419) to (1125, 591)\\
Label: Button, Text: Focus, right arrow, BoundingBox from (0, 593) to (1125, 764)\\
Label: Button, Text: Screen Time, right arrow, BoundingBox from (0, 766) to (1125, 938)\\
Label: Button, Text: settings, General, right arrow, BoundingBox from (0, 1046) to (1125, 1219)\\
Label: Button, Text: Control Center, right arrow, BoundingBox from (0, 1220) to (1125, 1393)\\
Label: Button, Text: Display \& Brightness, right arrow, BoundingBox from (0, 1395) to (1125, 1567)\\
Label: Button, Text: calendar, Home Screen, right arrow, BoundingBox from (0, 1568) to (1125, 1742)\\
Label: Button, Text: Accessibility, right arrow, BoundingBox from (0, 1742) to (1125, 1917)\\
Label: Button, Text: Wallpaper, right arrow, BoundingBox from (0, 1917) to (1125, 2089)\\
Label: Button, Text: Siri \& Search, right arrow, BoundingBox from (0, 2092) to (1125, 2266)\\
Label: Button, Text: smile, Face ID \& Passcode, right arrow, BoundingBox from (0, 2270) to (1125, 2423)\\
\textcolor{blue}{\textbf{Transition Action:}} \\ {Label: Button, Text: Display \& Brightness, right arrow, BoundingBox from (0, 1395) to (1125, 1567)} \\
\textcolor{blue}{\textbf{Caption of the Subsequent UI:}} \\
This UI screen is a settings page for display and brightness. It allows the user to adjust the appearance, brightness, text size, and other display settings of their device.
  \tcblower
\textcolor{blue}{\textbf{Predicting the Outcome of UI Actions}} \\
Question: What will the UI look like after tapping on the "Display \& Brightness" button? \\
Answer: The new UI screen might display options for adjusting the display appearance, including light and dark modes, as well as options for automatic brightness and true tone. \\
\textcolor{blue}{\textbf{Selecting a UI Element Given a Goal}} \\
Question: If I want to access the settings for adjusting the appearance and brightness, which option should I choose?  \\
Answer: To access the settings for adjusting the appearance and brightness, you should tap on the "Display \& Brightness" option.
\end{tcolorbox}
\vspace{-5mm}
\captionof{figure}{One example to illustrate the data generation for UI transition. }\label{fig:ui_transition}
\end{minipage}
\end{figure*}

\section{Vision Language Model}

To obtain the UI-focused instruction-tuned VLM, we employ the same network architecture as the LLaVA model~\cite{liu2023visual} with some modifications to the image input step. 
We found in initial experiments that the default CLIP encoder used by LLaVA used an input of 224x224 and resized input images to fit these dimensions. Resizing these images to such a small size and to fit the square shape causes significant loss of information and distortion, because of the relatively many intricate details of UI images compared to natural images. This in turn led to unsurprisingly poor interpretation of UI elements and text information. 

We made two modifications to address this issue. First, we pad the input UI images to square them before resizing to the required input size for the model, which prevents the UI images from becoming overly distorted. Second, we use CLIP-L-336px~\cite{radford2021learning} as the visual encoder, which increases the model input size to 336x336 pixels. We were concerned that this size would still be too small to capture UI detail, but our experiments showed that performance was much improved compared to the smaller image size.

Otherwise, our design matches LLaVA. The visual encoder is followed by the Vicuna-13B LLM~\cite{chiang2023vicuna, zheng2023judging} as a language decoder, 
as illustrated in Figure~\ref{fig:model}. Given the input UI image $\mathcal{I}$, we first pad and then resize the UI image to fit the visual encoder input. Grid features from before and after the last transformer layers of the pre-trained CLIP visual encoder are then used to output the visual features $h_v = \mathcal{E}(\mathcal{I})$. Following this, we apply a trainable feature alignment projection matrix $P$ to align the visual content into language embeddings $h_p = P\cdot h_v$, which maintain the same dimension as the word embedding in the language model. We then concatenate the projected visual embedding $h_p$ and the word embedding of the input language instruction $h_l$, and use the language decoder to generate the language response $R = \mathcal{D}(h_p, h_l)$.

\subsection{Data Preparation}

For each input UI image $\mathcal{I}$, and the generated pairs of questions and answers, $(Q_1, A_1, \ldots, Q_T, A_T)$, where $T$ represents the total number of turns, we arrange them into a sequential format. In this sequence, the user's instruction during the first turn is randomly selected from either $[Q_1, \mathcal{I}]$ or $[\mathcal{I}, Q_1]$, while for all subsequent turns, $Q_t$ is used as the user instruction. We consider all the corresponding answers, $A_t$, as the responses provided by the assistant at each respective turn.
This ensures that the sequence of user input, whether the UI image or language instruction comes first, does not affect the testing phase. 

\subsection{Fine-tuning Process}

\systemname uses the pretrained visual encoder $\mathcal{E}$ and the feature alignment projection matrix $P$ from LLaVA~\cite{liu2023visual} to align the visual content embedding with the pre-trained LLM word embedding.
We use the same language decoder as LLaVA $\mathcal{D}$, a pretrained Vicuna-13b-v1.3 LLM~\cite{zheng2023judging}.
We freeze the visual encoder weights while fine-tuning the feature alignment projection matrix and the LLM weights, resulting in a UI-focused instruction-tuned VLM.

\section{Evaluation}
We evaluate \systemname compared with the original LLaVA model in two ways. First, we analyze the model's ability to perform two basic UI Understanding tasks: identifying the existence of a UI element and a UI element's type. Second, we compare human preferences for UI screen descriptions generated by the two models. We plan further in-depth analyses as part of future work.

We ensure a fair comparison with the LLaVA model, we compare \systemname with a LLaVA model that uses the same 336-pixel-square vision encoder and language decoder. This LLaVA model is used for all comparisons.
For the evaluation, we sampled 100 UI images that had not been seen by \systemname previously during training or fine-tuning.
The full results of all evaluations are shown in Table~\ref{tbl:comparison}.

\def\arraystretch{1}%
\begin{table*}[t]
\scalebox{0.9}{\setlength\tabcolsep{5pt}
\begin{tabular}{lcccccc}
\hline
\multirow{ 2}{*}{\textbf{Agent}}  & \multicolumn{4}{|c|}{\textbf{Element Existence (\%)}} & \multicolumn{1}{|c|}{\textbf{Element Type (\%)} } & \multirow{ 2}{*}{\textbf{Human Preference (\%)}}   \\
   & \multicolumn{1}{|c}{\textbf{Accuracy}}  &  \textbf{Precision} &  \textbf{Recall}   & \multicolumn{1}{c|}{\textbf{F1 Score}}   & \multicolumn{1}{c|}{\textbf{Accuracy}}  &   \\
\hline
   \textbf{Ours}  & 67.94 & 67.32 & 70.23 & 68.74 & 25.67 & 72 \\
 LLaVA & 51.65 & 50.95 & 99.59 & 67.41 & 9.03 & 20 \\
\bottomrule
\end{tabular}
}
\caption{Comparison between \systemname and LLaVA on all evaluation tasks. }
\label{tbl:comparison}
\end{table*}

\subsection{UI Understanding Tasks}

We use the same UI element detection model~\cite{zhang2021screen} used earlier for data generation to produce data that we take as ground truth for evaluating models' UI understanding capabilities. The model provides bounding boxes and element types for all the UI elements in each evaluation UI image. 

In order to evaluate UI element detection, for each screen in the evaluation set we produce a set of positive and negative detection samples. 5 positive samples are taken from the elements detected on each screen. 5 negative examples are randomly sampled from detections on other screens in the evaluation set, but only after a basic matching check is applied to ensure an equivalent element is not likely to appear on the original screen (e.g., close and back icons appear relatively commonly). We query both models for each of the 5 positive and 5 negative samples, asking whether they exist in the input UI image. Our \systemname agent achieves an accuracy of 67.94\%, outperforming the LLaVA model, which has a 51.65\% accuracy. Note that the LLaVA model nearly always provides positive responses, resulting in a high recall of 99.59\% but close to random-chance accuracy.
 
In order to evaluate UI element type, we randomly sample 5 UI elements from each UI and query both models to identify the types of these UI elements from a list of 12 common types, including button, checkbox, container, dialog, icon, page control, picture, segmented control, slider, text, text field, and toggle. The list of element types is provided in a randomized order for each query. This is a harder task, and our \systemname agent achieves an accuracy of 25.67\% compared to 9.03\% for LLaVA.

\subsection{Human Evaluation of UI Descriptions}

For all 100 images in the evaluation set, we queried both models to produce a UI description. We then built a rating user interface in a webpage, where the screen image and descriptions from both models were displayed. The descriptions from each model were shown in a randomized order on the page, so a human rater could not predict which model generated which description. Raters were asked which description they preferred or if both descriptions were about the same. All authors participated as raters for this evaluation. The results show that 72\% of the responses preferred the \systemname generated description, in comparison to 20\% that preferred LLaVA's description. The remaining 8\% were rated similar.

\begin{table*}
\begin{center}
\begin{tabular}[t]{|l p{14.4cm}|} 
 \hline
 & 
\includegraphics[width=0.15\textwidth]{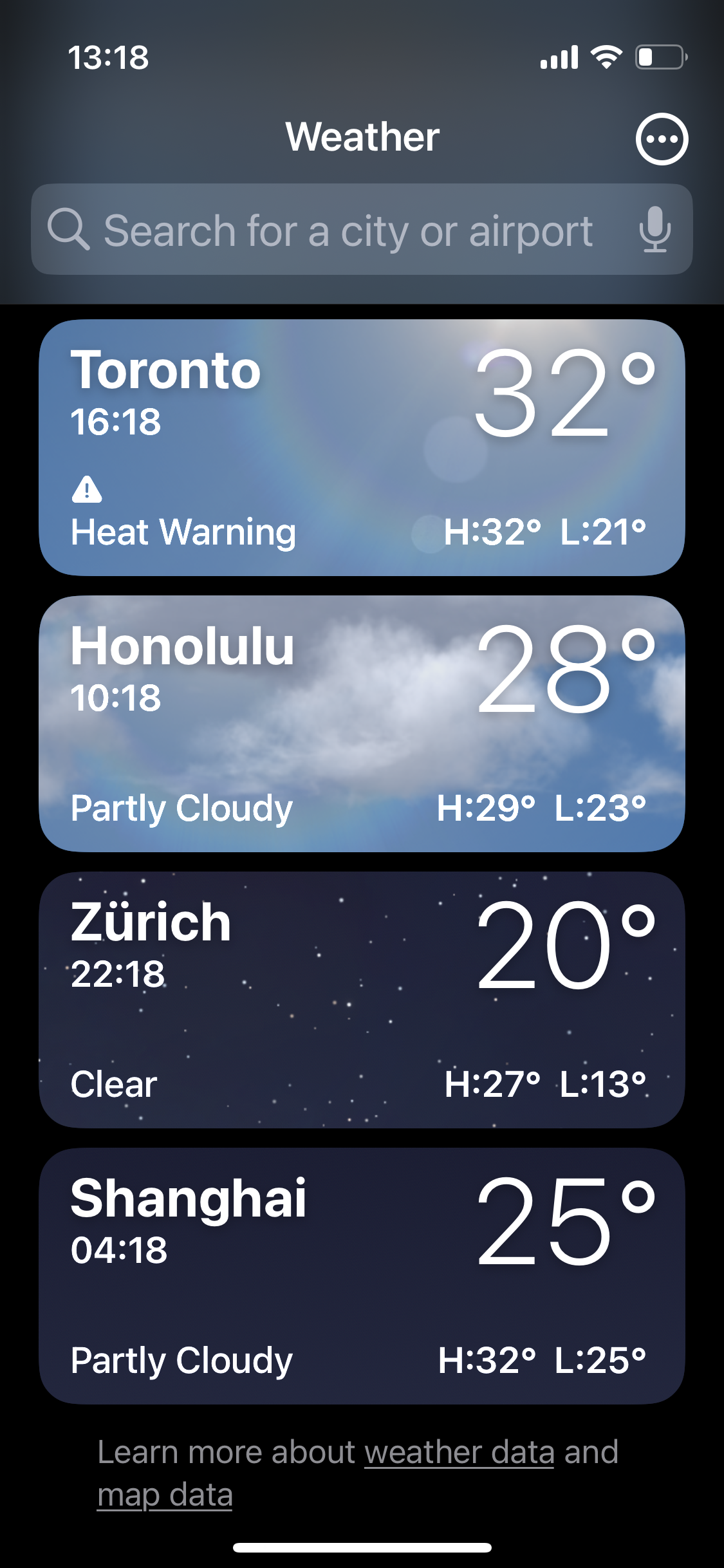}
 \\ 
 \hline
  Type & \textbf{Detailed Description} \\ 
 \hdashline
 User & Describe the UI. \\ 
 \hdashline
 Our Response & 
This UI screen presents weather information for various cities. At the top, the title "Weather" is centrally positioned, and to the right, a settings icon is available for users to adjust their preferences. Below the title, a search field allows users to look for weather information in a specific city or airport.
The main part of the screen is occupied by a series of buttons, each representing a different city. Each button includes the city name, the current time, temperature, weather condition, and the high and low temperatures for the day. The cities listed are Toronto, Honolulu, Zurich, and Shanghai. Notably, the city of Toronto is accompanied by a warning for "Heat Warnings".
At the bottom of the screen, a text section provides additional information about weather data, encouraging users to learn more about weather data and how to interpret it.  \\
\hdashline
 LLaVA Response & The UI of the iPhone screen features a weather app with a variety of weather information displayed. The app provides temperature readings for different cities, including Toronto, Zurich, and Shanghai, as well as the current time. The app also displays a map with the current weather conditions for each city. The UI is designed to be user-friendly and easy to navigate, allowing users to quickly access and view the weather information they need.  \\
 \hline
   Type & \textbf{Details on UI} \\ 
 \hdashline
 User & What is the weather condition of Honolulu? \\
  \hdashline
 Our Response &   The weather condition of Honolulu is "Partly Cloudy".  \\
 \hdashline
 LLaVA Response &  The weather condition of Honolulu is sunny.  \\
  \hline
    Type & \textbf{Logical Reasoning} \\ 
 \hdashline
 User & If I want to know whether it is raining in London, what should I do? \\ 
 \hdashline
 Our Response & To determine if it is raining in London, you should enter "London" in the search bar. The weather information for London will be displayed, including the current weather conditions. If it is raining, the relevant weather conditions (e.g., ``Raining'') will be shown. \\
\hdashline
 LLaVA Response & To find out whether it is raining in London, you can check the weather app on your iPhone. Look for the city of London in the list of cities displayed on the app, and check the weather forecast for that day. The app should provide you with the current weather conditions and a forecast for the upcoming days, including whether it is raining or not. If it shows rain for London, then it is likely to rain in London that day.  \\
 \hline
   Type & \textbf{UI Transition} \\ 
 \hdashline
 User & If I click on the top right button, what will happen? \\
  \hdashline
 Our Response & If you click on the top right button, it is likely that the settings for the weather information will be shown. This could include changing the city for which the weather information is displayed, adjusting the units of measurement (e.g., Celsius or Fahrenheit), and other settings related to the weather information. \\
 \hdashline
 LLaVA Response &  If you click on the top right button on the weather app, it will likely open the current weather conditions for the selected city. The button is labeled "Today's weather," which suggests that it will display the current weather information for the city. This could include temperature, humidity, wind speed, and precipitation, among other relevant weather data. \\
  \hline
\end{tabular}
\end{center}
\label{tbl:result1}
\caption{Example prompt demonstrating our agent's capability compared to the original LLaVA agent. }
\end{table*}

\begin{table*}[!]
\begin{center}
\begin{tabular}[t]{|l p{14.4cm}|} 
 \hline
 & 
\includegraphics[width=0.15\textwidth]{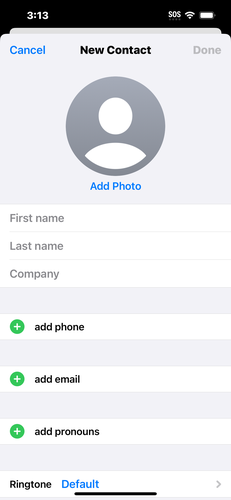}
 \\ 
 \hline
  Type & \textbf{Detailed Description} \\ 
 \hdashline
 User & Could you provide a detailed description? \\ 
 \hdashline
 Our Response & 
This UI screen is designed for creating a new contact. At the top of the screen, there are three options: ``Cancel'' on the left, ``New Contact'' in the center, and "Done" on the right. These options allow the user to either discard the process, understand the current operation, or complete the contact creation, respectively.
In the middle of the screen, there's a section labeled ``Add Photo'', which suggests the ability to add a photo to the new contact. Below this, there are fields for entering the contact's first and last names. The ``First name'' field is located above the ``Last name'' field.
Further down, there's a section for entering the contact's company. This section is followed by three buttons for adding various types of contact information. The first button, labeled ``Add phone'', allows the user to add a phone number. The second button, labeled ``Add email'', provides an option to add an email address. The third button, labeled ``Add pronouns'', offers the ability to add any additional pronouns associated with the contact.
At the bottom of the screen, there's a button labeled ``Ringtone Default'', which presumably allows the user to set a default ringtone for this contact. Next to this button, there's a right arrow icon, suggesting that there might be additional options or information available upon interaction.  \\
\hdashline
 LLaVA Response & The image shows a close-up view of a smartphone screen displaying a contact list. The contact list is divided into two sections: one section contains the names of contacts, and the other section contains their phone numbers. The contact list is organized in a grid-like layout, with each contact's name and phone number displayed in separate rows. This layout makes it easy for the user to quickly identify and access the contact information they need. The phone number section is highlighted in green, which helps to distinguish it from the contact names section. Overall, the contact list is a useful feature on the smartphone that allows users to easily manage and access their contacts' information.  \\
 \hline
    Type & \textbf{Logical Reasoning} \\ 
 \hdashline
 User & what should I do if I want to add Amy at Company ABC with email address 12345@abc.com and have the Circuit Ringtone when she gives a phone call to me?
 \\ 
 \hdashline
 Our Response & To add a contact named Amy from Company ABC with the email address 12345@abc.com and set the Circuit Ringtone, you should first enter "Amy" in the "First name" text field and "Company ABC" in the "Company" text field. Then, in the "Email" section, you should enter "12345@abc.com". Scroll down to the "Ringtone" section and click on the "Default" button to set the Circuit Ringtone as the default ringtone for Amy's contact. After setting the ringtone, you can confirm the creation of the contact by clicking on the "Done" button located at the top right of the screen. \\
\hdashline
 LLaVA Response & To add Amy at Company ABC with email address 12345@abc.com and have the Circuit Ringtone when she gives a phone call to you, you should follow these steps:
Open the phone app on your device.
Navigate to the contact list or address book.
Tap the "Add Contact" or "Add New Contact" button.
Enter her name, email address (12345@abc.com), and any other relevant information you would like to include.
Save the contact by tapping the "Save" or "Add" button.
Set the Circuit Ringtone as her ringtone by navigating to the phone settings, selecting "Ringtones," and choosing the Circuit Ringtone.
Save the changes to the phone settings.  \\
  \hline
\end{tabular}
\end{center}
\label{tbl:result2}
\caption{Example prompt demonstrating our agent's capability compared to the original LLaVA agent. }
\end{table*}

\section{Discussion and Future Work} 
This paper presents an initial promising step towards producing a Vision-Language Model that can perform general UI understanding tasks. Our approach adapts an existing method for the UI domain, and shows that it is able to produce UI-centric instruction-tuned behavior despite being trained with synthetic textual data produced from a large UI dataset of only screenshots. 

There are many potential applications for such a model, including providing accessible descriptions of screens to vision-impaired users, repairing automated UI tests, or providing contextual help to assist a user with navigating an interface. We have informally completed some of these tasks successfully, and a basic example of our model being used for UI navigation can be seen in Table 4.%

More comprehensive evaluation will be needed to fully understand the performance of this model and its progeny. An open question is: what sort of tasks or benchmarks should be adopted to evaluate the performance of a model on UI tasks? Here we use very basic UI understanding tasks as initial benchmarks, including UI element detection and element type identification.
There are further simple evaluations of this form, such as icon recognition. There are also other higher-level tasks along which a model could be evaluated. Spotlight~\cite{Li2022SpotlightMU} suggests some tasks, including widget captioning, screen summarization, command grounding, and tappability prediction. A goal for the research community should be to work towards agreement on a standard set of benchmarks that can be automatically and easily evaluated when a new models is produced, which might help ensure that others working on models outside of the HCI community produce models that can successfully perform UI understanding tasks.

Other future research could delve deeper into refining the model's reasoning abilities, expanding its knowledge base for UI design, developing novel agents to facilitate effective UI interaction and navigation, and ensuring robustness across broader UI variations. 

Future work can focus on the following aspects to further develop the agent:
\begin{itemize}
    \item Improving dataset quality: While our generated UI dataset has helped in training our \systemname~ agent, there is still room for improving the quality and diversity of the dataset. A deeper analysis of our generated dataset is needed to better understand its flaws, which can lead to improved prompting strategies. Future work on synthetic data could focus on reducing hallucination, adding more annotated data, and varying UIs from different domains. Human-annotated data could also be added, especially to assist with common-sense reasoning and task planning. 
    \item Improving the model: The performance of \systemname~ could be improved by building on top of more advanced vision-language model architectures. In particular, it would be ideal to build upon an image encoder that can function with a variety of input resolutions, including very high resolutions. Such a model could them more easily work on traditional desktop user interfaces, for example.
    \item Support for tools and machine-interpretable output: Currently, \systemname seems to be fairly limited in its ability to produce structured outputs that could be more easily interpreted by a machine, such as JSON. Some applications, particularly UI navigation and software testing, will require the models to work more easily with other traditional software, so additional training and fine-tuning will be required to make this possible and reliable.
\end{itemize}

\begin{table*}[!]
\begin{center}
\begin{tabular}[t]{|l|} 
 \hline
  Task: start a stopwatch. \\ 
 \hdashline
 \begin{minipage}[t]{0.3\linewidth}
    \vspace*{-2pt}
    \includegraphics[width=1\textwidth]{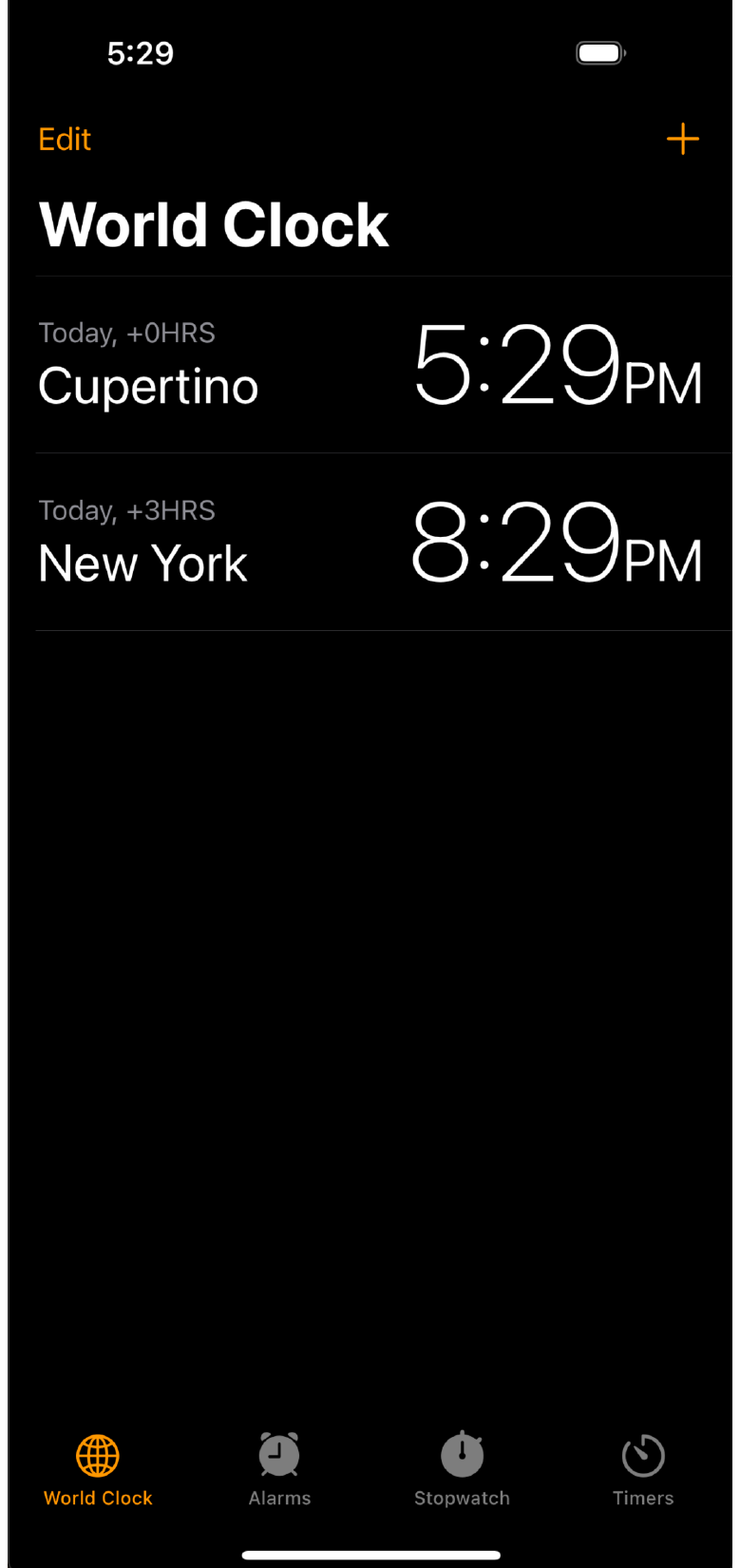}
    \end{minipage}%
    \begin{minipage}[t]{0.6\linewidth}
    \vspace*{8pt}
    {Human: Describe the screen and its key features, which may be useful for performing various actions on the screen.\\ 
    \\
    \systemname: [Description of the UI] \\ \\
    Human: What action (tap, swipe, enter text) should a user perform on this screen to do the following: start a stopwatch. \\ \\
    \systemname: To start a stopwatch on this screen, a user should tap on the "Stopwatch" tab located in the middle of the UI screen. This tab is positioned between the "Currency" and "Settings" tabs. By tapping on the "Stopwatch" tab, the user can initiate the stopwatch functionality and start timing events.}
    \end{minipage}%
   \\
\hdashline
\begin{minipage}[t]{0.3\linewidth}
    \vspace*{-2pt}
    \includegraphics[width=1\textwidth]{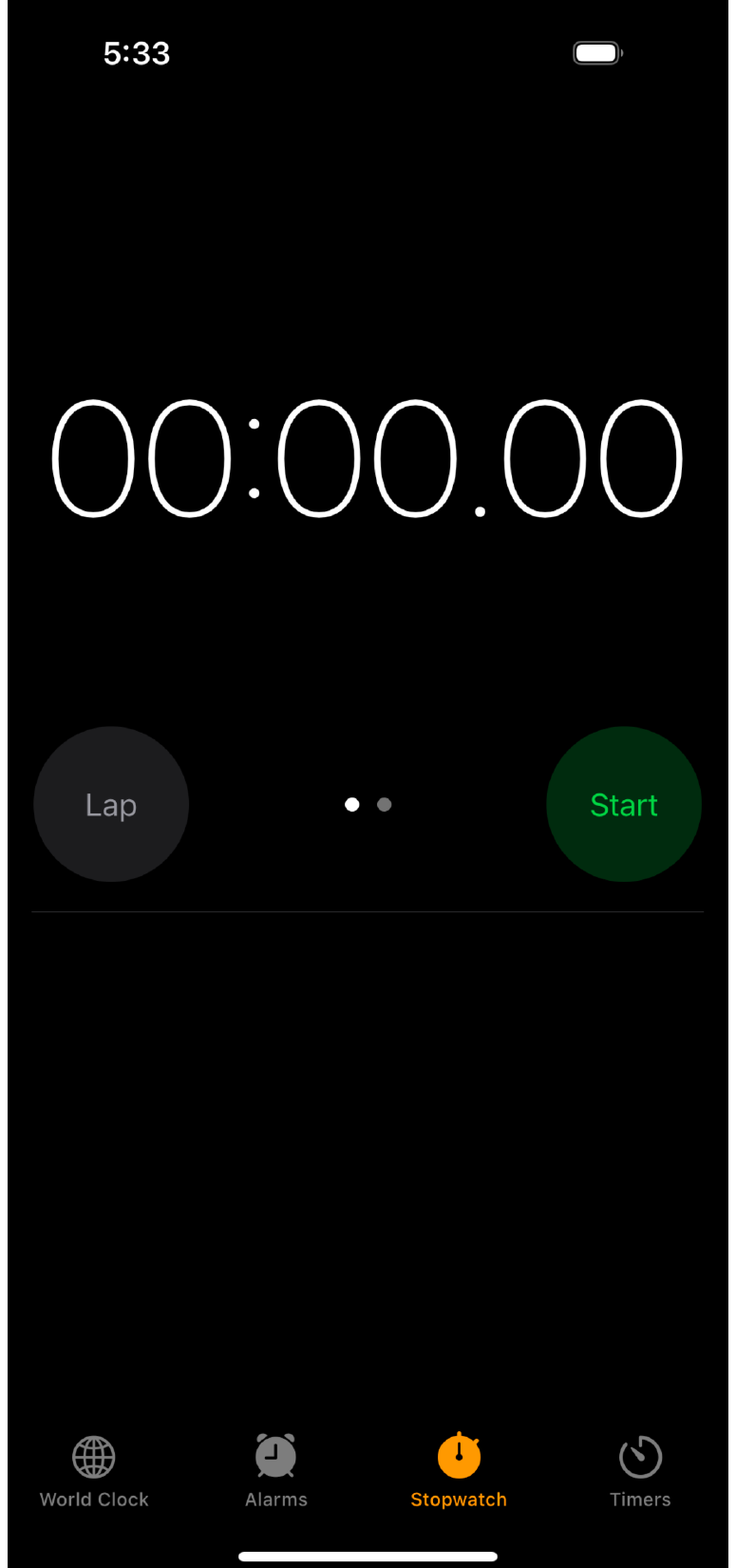}
    \end{minipage}%
\begin{minipage}[t]{0.6\linewidth}
    \vspace*{23pt}
    {Human: Describe the screen and its key features, which may be useful for performing various actions on the screen.\\ 
    \\
    \systemname: [Description of the UI] \\ \\
    Human: What action (tap, swipe, enter text) should a user perform on this screen to do the following: start a stopwatch. \\ \\
    \systemname: To start a stopwatch, the user should tap on the "Start" button located below the timer.}
    \end{minipage}%
 \\ 
  \hline
\end{tabular}
\end{center}
\label{tbl:navigation}
\caption{Example demonstrating our agent's potential capability of planning UI navigation. }
\end{table*}

\section{Conclusion} 

In this paper we have shown how to address a significant gap in the current abilities of Vision-Language Models to understand UIs. We proposed a data generation and fine-tuning approach that can adapt VLMs to the realm of UIs. 
The challenges posed by UIs go beyond static elements and encompass interactive components that introduce both structural and functional complexity. Our approach, building on the data generation and training of LLaVA~\cite{liu2023visual}, showcases its versatility in capturing diverse responses relevant to UI understanding and interactions, ranging from individual UI element properties to elaborate UI descriptions, reasoning, potential user actions, and dynamic UI transitions.
The performance evaluation of our UI-focused instruction-following visual agent highlights its effectiveness in comparison to existing models that are not trained on UI data.

\bibliographystyle{ACM-Reference-Format}
\bibliography{Reference}

\end{document}